\documentclass[prb,reprint,superscriptaddress,amsmath,amssymb,floatfix,twocolumn]{revtex4-1}

\usepackage{graphicx}
\usepackage{wrapfig}
\usepackage{enumerate}

\begin{document}
\title{Correlated electronic structure with uncorrelated disorder}
\author{A. \"Ostlin}
\affiliation{Theoretical Physics III, Center for Electronic
Correlations and Magnetism, Institute of Physics, University of
Augsburg, D-86135 Augsburg, Germany}
\author{L. Vitos}
\affiliation{Department of Materials Science and Engineering, Applied Materials Physics,
KTH Royal Institute of Technology, SE-10044 Stockholm, Sweden}
\affiliation{Department of Physics and Astronomy, Division of Materials Theory,
Uppsala University, Box 516, SE-75120 Uppsala, Sweden}
\affiliation{Research Institute for Solid State Physics and Optics, Wigner Research Center for Physics, P.O. Box 49, H-1525 Budapest, Hungary}
\author{L. Chioncel}
\affiliation{Augsburg Center for Innovative Technologies, University of Augsburg,
D-86135 Augsburg, Germany}
\affiliation{Theoretical Physics III, Center for Electronic
Correlations and Magnetism, Institute of Physics, University of
Augsburg, D-86135 Augsburg, Germany}
 
\begin{abstract} 
We introduce a computational scheme for calculating the 
electronic structure of random alloys that includes electronic correlations within the framework of the combined density functional and dynamical mean-field theory. By making use of the particularly simple parameterization of the
electron Green's function within the linearized muffin-tin orbitals method, we show that it is possible to greatly
simplify the embedding of the self-energy. This in turn facilitates the implementation of the
coherent potential approximation, which is used to model the substitutional disorder. 
The computational technique is tested on the Cu-Pd binary alloy system, and
for disordered Mn-Ni interchange in the half-metallic NiMnSb.
\end{abstract}

\maketitle 
\section{Introduction}\label{sec_intro}

Disordered metallic alloys find applications in a large number of
areas of materials science. 
Designing alloys with specific thermal, electrical, and mechanical properties such as
conductivity, ductility, and strength, nowadays commonly starts at the microscopic level~\cite{olson.00,ja.on.13,vi.ko.02}.
First-principles calculations of the electronic structure offer
a parameter-free framework to meet specific engineering demands for materials prediction.
Advances in the atomistic simulation of physical properties should to a large extent be attributed
to the development of density functional theory (DFT)~\cite{ho.ko.64,ko.sh.65} within
the local density approximation (LDA) or beyond-LDA schemes.

In solids exhibiting disorder the calculation of any physical property involves
configurational averaging over all realizations of the random variables
characterizing the disorder. In the case of substitutional disorder the symmetry  of the lattice is kept,
but the type of atoms
in the basis is randomly distributed.
This causes the crystal to lose translational symmetry, making the Bloch
theorem inapplicable. 
Perhaps the most successful approach to solve the problems associated with substitutional disorder
is the coherent potential approximation~\cite{soven.67} (CPA).
The presence of random atomic substitution generates a fluctuating
external potential which within the CPA is substituted by an effective medium. 
This effective medium is energy dependent and is determined self-consistently through
the condition that the impurity scattering should vanish on average. The CPA is
a single-site approximation, which becomes exact in certain limiting cases~\cite{ve.ki.68}.
The explanation for the 
good accuracy of the CPA can be traced back to the fact that it 
becomes exact in the limit of
infinite lattice coordination number, $Z \rightarrow \infty$~\cite{vl.vo.92}.

The CPA was initially applied to tight-binding Hamiltonians~\cite{ve.ki.68}, 
where for binary alloys the CPA equations take a complex polynomial form that
can be solved directly. Later Gy\H{o}rffy~\cite{gyorffy.72} formulated
the CPA equations for the muffin-tin potentials within the multiple-scattering
Korringa-Kohn-Rostocker (KKR) method~\cite{korr.47,ko.ro.54}.
Consequently, the configurational average could be performed over the
scattering path operator, instead of the Green's function, simplifying the 
implementation of the CPA for materials calculations~\cite{st.wi.71}.
Later the CPA was also implemented within the linearized muffin-tin
orbitals~\cite{ande.75} (LMTO) basis
set~\cite{ku.dr.87,ku.dr.90,ab.ve.91,ab.sk.93}. With the advent of the third-generation exact 
muffin-tin orbitals~\cite{an.je.94.2,vito.01,vitos.10} (EMTO) method, and the 
full-charge density~\cite{vi.ko.94} (FCD) technique, it was possible to go beyond
the atomic-sphere approximation (ASA) with CPA calculations~\cite{vi.ab.01}, and
investigate the energetics of anisotropic lattice distortions.
For interacting model Hamiltonians, dynamical mean-field theory~\cite{me.vo.89,ge.ko.96,ko.vo.04} (DMFT)
was also combined with the CPA~\cite{ja.vo.92,ul.ja.95,kake.02}.
Later on the methodology was extended to study realistic materials 
containing significant electronic correlations within the
framework of a combined DFT+DMFT method~\cite{ko.sa.06,held.07}.
To treat weak disorder within the framework of charge
self-consistent LDA+DMFT, the CPA has been implemented within the KKR method~\cite{mi.ch.05}.
In an alternative approach, the band structures computed from DFT were mapped to
tight-binding model Hamiltonians were the disorder was treated within the CPA~\cite{wi.to.11,ko.pc.14,be.po.15,be.an.16}.

In this paper, we introduce a method that can treat substitutional disorder effects through the
CPA, and electronic correlation effects through DMFT, on an equal footing
for real materials. 
The method is based on the LDA+DMFT method, $z$MTO+DMFT, which was recently introduced by us~\cite{os.vi.17}.
By making use of the particularly simple parameterized form of the electron Green's function
in a linearized basis set, we show that the self-energy can be incorporated naturally within
the LMTO formalism as a modification of a single, self-consistently determined parameter.
Due to this straightforward inclusion of the many-body effects, the CPA
within the LMTO method retains its general form, and can be used almost unaltered
for the LDA+DMFT approach.

The paper is organized as follows: Section~\ref{sec_muffintin} briefly reviews the muffin-tin basis sets used in this work.
A short overview of the LMTO method is given in Sec.~\ref{sec_lmto}, with additional
formulas given in Appendix~\ref{app_lmto}, in order to introduce the most important
quantities needed for the CPA+DMFT implementation.
A review of the EMTO method, which is the second basis set used in this paper, is given in Appendix~\ref{app_emto}.
In Sec.~\ref{sec_corrdis} we present the most important development in this paper, a combination
of the CPA and the LDA+DMFT method.
First, in Sec.~\ref{sec_avg} we discuss the principles behind the configuration average
used for the CPA. Sec.~\ref{sec_modparam} shows how the electronic self-energy can be
incorporated into the LMTO potential functions, while Sec.~\ref{sec_cpaloop} demonstrates
how the potential functions are used as an effective medium within the CPA. The section
ends with an outline of the full computational scheme. 
In Sec.~\ref{sec_res}
we present the results of the method applied to the binary copper-palladium system,
treating the $d$-electrons of palladium as correlated. The half-metallic semi-Heusler compound
NiMnSb with disorder is also investigated. Section~\ref{sec_conc} provides
a conclusion of our paper.

\section{Electronic structure with muffin-tin orbitals}\label{sec_muffintin}

The standard approach adopted in first-principles electronic structure calculations
is the mapping to an effective single-particle equation.
The Hohenberg-Kohn-Sham density functional formalism~\cite{ho.ko.64,ko.sh.65} provides a
self-consistent description for the effective one-particle potential $V_{\mathrm{eff}}({\bf r})$.
Within the muffin-tin orbital methods, the effective potential in the Kohn-Sham equations,
\begin{equation}\label{kosh}
\left[\nabla^2-V_{\mathrm{eff}}({\bf r})\right] \Psi_{j}({\bf r})
\;=\;\epsilon_j\Psi_j({\bf r}),
\end{equation}
is approximated by spherical potential wells $V_R(r_R)-V_0$
centered on lattice sites $\textbf{R}$, and a constant interstitial potential
$V_0$, viz.,
\begin{equation}
V_{\mathrm{eff}}(\textbf{r}) \approx V_{mt}(r_R) \equiv V_0 + \sum\limits_{R}
[V_R(r_R)-V_0],
\end{equation}
where we introduced the notation $\textbf{r}_R \equiv r_R \hat{r}_R=\textbf{r}-\textbf{R}$.
We will in the following omit the vector notation for simplicity. This form
of the potential makes it possible to divide the Kohn-Sham equation~(\ref{kosh})
into radial Schr\"odinger-like equations within the muffin-tin spheres, and wave equations
in the interstitial region, which can be solved separately.
The computational scheme that we present in this paper is based on the muffin-tin theories
developed by Andersen and coworkers, namely the LMTO~\cite{oka.70,an.je.84,an.je.86,skri.84}, and the EMTO~\cite{an.je.94.2,an.sa.00,vito.01,vi.sk.00,vitos.10}, methods. In the following, we briefly 
review the main ideas and notations behind the LMTO-ASA method.

\subsection{Linear Muffin-Tin Orbitals}\label{sec_lmto}

The \emph{energy-independent} linearized muffin-tin orbitals $\chi^\alpha_{RL}(r_R)$, centered at the lattice site $R$, are given as:
\begin{eqnarray}
\chi^\alpha_{RL}({\bf r}_R) &=& \phi_{RL}({\bf r}_R) + \sum_{R'L'} \dot{\phi}^\alpha_{R'L'}({\bf r}_R)h^\alpha_{RL,R'L'}, \\
\dot{\phi}^\alpha_{RL}({\bf r}_R) &=& \dot{\phi}_{RL}({\bf r}_R) + \phi_{RL}({\bf r}_R) o^\alpha_{RL}, \\
\phi_{RL}({\bf r}_R) &=& \phi_{RL}(r_R) Y_L(\hat{r}_R).
\end{eqnarray}
$\phi_{RL}(r_R)$ is the solution of the radial Schr\"odinger equation at an arbitrary energy $\epsilon_\nu$, usually taken to be the center of gravity of the occupied part of the band, and $o^{\alpha}_{RL} = \langle \phi_{RL} \vert \dot{\phi}^\alpha_{RL} \rangle$ is an overlap integral. $L\equiv(l,m)$ denotes the orbital and azimuthal quantum numbers, respectively. The superscript $\alpha$ denotes the screening representation used in the tight-binding LMTO theory. The expansion coefficients $h^\alpha$ are determined from the condition that the wave function is continuous and differentiable at the sphere boundary at each sphere:
$h^\alpha_{RL,R'L'}({\bf k})  
=(c^\alpha_{RL} - \epsilon_\nu) \delta_{RR'} \delta_{LL'} + 
 \sqrt{d^\alpha_{Rl}}S^{\alpha}_{RLR'L'}({\bf k})\sqrt{d^\alpha_{R'l'}}$.
Note that we are now assuming a translationally invariant lattice system, so
that the Bloch wave vector \textbf{k} is well defined. From now on, $R$ will
denote the atoms in the unit cell only. 
The coefficient $h^\alpha_{RL,R'L'}$ is parametrized by the center of the band,
$c^\alpha_{RL}= \epsilon_\nu - P^\alpha_{RL}(\epsilon_\nu) [\dot{P}^\alpha_{RL}(\epsilon_\nu)]^{-1}$, and 
the band width parameter $d^\alpha_{RL} = [\dot{P}^\alpha_{RL}(\epsilon_\nu)]^{-1}$, both expressed
in terms of the potential function $P^\alpha_{RL}$ and its energy derivative $\dot{P}^\alpha_{RL}$ evaluated at $\epsilon_\nu$.
The potential function $P^{\alpha}$ and the structure constant $S^{\alpha}$ are expressed using the conventional potential function $P^0$ and 
the conventional structure constant matrix $S^0$~\cite{an.je.86}:
\begin{eqnarray}\label{transform}
P^\alpha_{RL} &=& [P^0(1-\alpha P^0)^{-1}]_{RL},\nonumber\\
S^\alpha_{RLR'L'} &=& [S^0(1-\alpha S^0)^{-1}]_{RLR'L'}.
\end{eqnarray}
$P^0_{RL}$ is proportional to the cotangent of the phase shift created by the potential centered at the sphere at $R$. 
Thus, the potential parameters $P$
characterize the scattering properties of the atoms placed at the lattice sites.
The geometry of the lattice enters through the structure constants $S^0$, which is independent
of the type of atoms occupying the sites. $S^0$ has a long range behavior in real space,
but is decaying nearly exponentially in the tight-binding $\beta$-representation~\cite{an.je.84}.
The so-called band distortion parameter $\alpha = (P^0)^{-1} -(P^\alpha)^{-1} = (S^0)^{-1} -(S^\alpha)^{-1}$, which
is also used to denote the screening representation, gives a
relation between the $\alpha$-representation and the unscreened representation.
In the nearly-orthogonal $\gamma$-representation, $\alpha=\gamma$, and $o^\gamma_{RL}=0$, hence
the Hamiltonian $H^{\gamma}_{RLR'L'}$ is given by:
\begin{equation}\label{hamil}
H^{\gamma}_{RLR'L'}({\bf k})=C_{Rl}+\sqrt{\Delta_{Rl}}S^{\gamma}_{RLR'L'}({\bf k})\sqrt{\Delta_{R'l'}},
\end{equation}
where $C_{Rl}$, $\Delta_{Rl}$ and $\gamma_{Rl}$ are the representation-independent band center-,
width- and distortion potential parameters, respectively~\cite{an.je.86,skri.84}. 
The corresponding Green's function in the $\gamma$-representation is given by:
\begin{equation}\label{green}
G^{\gamma}_{RLR'L'}({\bf k},z) = [\left(z-H^{\gamma}({\bf k}))\right ]^{-1}_{RLR'L'},
\end{equation}
where $z$ is an arbitrary complex energy.
Further important relations among the LMTO representations can be found in Appendix~\ref{app_lmto}.

\section{Electronic correlations and disorder: the single-site approximation}
\label{sec_corrdis}
In this section we present an LMTO-CPA scheme, that allows to include local self-energies, on the level of DMFT, for the alloy components. The scheme is implemented within the Matsubara representation,
and is combined with the $z$MTO+DMFT method~\cite{os.vi.17}. 
Section~\ref{sec_avg} briefly discusses the configurational averaging and the CPA, while in Sec.~\ref{sec_modparam} we show how DMFT through the Dyson equation leads to a renormalization of the parameters of the LMTO-ASA formalism. 
Section~\ref{sec_cpaloop} combines the ideas of the previous two sections, and proposes a combined CPA and DMFT loop.

\subsection{Configuration averaging and the CPA}\label{sec_avg}

In disordered systems the configurational degrees of freedom characterizing the composition are described by a random variable. Consequently, the potentials at sites are random in space as in quenched disordered solids. A particular realization of the random variable constitutes a configuration of the system in discussion. 
According to Anderson~\cite{ande.58} only physically measurable quantities such as diffusion probabilities, response functions, and densities of states should be configurationally averaged. As these quantities are themselves Green's functions,  electronic structure methods using Green's functions are favored for the study of disordered systems.

A major development of the theory of disordered electronic systems was achieved using the CPA. The CPA belongs to the class of mean-field theories
according to which the properties of the entire material are determined from the average behavior at a subsystem, usually taken to be a single site (cell) in the material. 
In the multiple scattering description of a disordered system, one considers the propagation of an electron through a disordered medium as a succession of elementary scatterings at the random atomic point scatterers. 
In the single-site approximation one considers only the independent scattering off different sites and finally takes the average over all configurations of the disordered system consisting of these scatterers. 
One may then consider any single site in a specific configuration and replace the surrounding material by a translationally invariant medium, constructed to reflect the ensemble average over all configurations. 
In the CPA this medium is chosen in a self-consistent way. One assumes that averages over the occupation of a site embedded in the effective medium yield quantities indistinguishable from those associated with a site of the medium itself. 

In view of the great progress achieved through the previous implementations of the CPA within muffin-tin orbitals methods~\cite{st.wi.71,gyorffy.72,ku.dr.87,ku.dr.90,ab.ve.91,ab.sk.93}
we present here in detail a novel combination of CPA+DMFT in the 
recently developed $z$MTO+DMFT method~\cite{os.vi.17}.

\subsection{Self-energy-modified effective potential parameters}\label{sec_modparam}

To deal with the important question concerning the effect of interaction, we start by observing that within the DMFT the self-energy is local and primarily modifies the local parameters of the model that describes disorder. 
In this section, we show that the presence of a local self-energy, $\Sigma_{RLRL'}(z)$, modifies the potential function $P$ entering in the expression of the Green's function in the LMTO-ASA formalism. 

We start from the Dyson equation used to construct the LDA+DMFT Green's function:
\begin{equation}\label{dmftdyson}
[G^{\mathrm{DMFT}}_{RLR'L'}({\bf k},z)]^{-1}=[G^{\mathrm{LDA}}_{RLR'L'}({\bf k},z)]^{-1}-\Sigma_{RLRL'}(z),
\end{equation}
where $G^{\mathrm{DMFT/LDA}}_{RLR'L'}$ denotes the LDA+DMFT/LDA-level Green's function and $\Sigma(z)$ the self-energy.
It is useful to define an auxiliary Green's function, the path operator $g^{\alpha}_{RLR'L'}$, as
\begin{align}\label{pathop}
g^{\alpha}_{RLR'L'}({\bf k},z)=\left[P^{\alpha}_{Rl}(z)-S^{\alpha}_{RLR'L'}({\bf k})\right]^{-1},
\end{align}
which is valid for a general representation $\alpha$. In Appendix~\ref{app_lmto}
we present explicit expressions for the potential functions and auxiliary Green's functions
in different representations, as well as their connection to the physical Green's function.
As is apparent in Eq.~(\ref{pathop}), the full energy- and \textbf{k}-dependence of the path operator
is contained in the potential function and the structure constants, respectively. Furthermore, the
potential function is fully local, i.e., it is diagonal in site index.
In the following, it will prove convenient to first work in the $\gamma$-representation, since here
the Green's function takes a particularly simple form (see Eq.~(\ref{gammagreen})):
\begin{equation}\label{pathtog}
G^{\gamma,\mathrm{LDA}}_{RLR'L'} ({\bf k},z) = \Delta^{-1/2}_{Rl}g^{\gamma}_{RLR'L'}({\bf k},z)\Delta^{-1/2}_{R'l'},
\end{equation}
i.e., it is the path operator normalized by the potential parameter $\Delta_{Rl}$.
Using Eqs.~(\ref{pathtog}) and~(\ref{potf}), the LDA Green's function can be evaluated as follows:
\begin{align}
[G^{\gamma,\mathrm{LDA}}_{RLR'L'} ({\bf k},z)]^{-1} 
&= \Delta^{1/2}_{Rl} \left[ P^{\gamma,\mathrm{LDA}}_{Rl}(z)-S^{\gamma}_{RLR'L'}({\bf k}) \right]  \Delta^{1/2}_{R'l'} \nonumber\\
&= z-C_{Rl}-\Delta^{1/2}_{Rl}S^{\gamma}_{RLR'L'}({\bf k}) \Delta^{1/2}_{R'l'}.
\end{align}
Hence, the LDA+DMFT Green's function~(\ref{dmftdyson}) can be written as
\begin{multline}\label{dmftdyson2}
[G^{\gamma,\mathrm{DMFT}}_{RLR'L'}({\bf k},z)]^{-1}=\\z-C_{Rl}-\Delta^{1/2}_{Rl}S^{\gamma}_{RLR'L'}({\bf k})\Delta^{1/2}_{R'l'}-\Sigma_{RLRL'}(z),
\end{multline}
i.e., as the resolvent of the Hamiltonian~(\ref{hamil}) with an embedded self-energy.
From Eq.~(\ref{dmftdyson2}), it is obvious that the same result will
follow if the potential parameter $C_{Rl}$ is replaced by an effective parameter, in which the
self-energy is embedded, viz.,
\begin{equation}
C^{\mathrm{DMFT}}_{RLRL'}(z) \equiv C_{Rl}+\Sigma_{RLRL'}(z).
\end{equation}
Hence the effective potential parameter $C^{\mathrm{DMFT}}_{RLRL'}$ will now in general be complex, energy-dependent, and have off-diagonal elements. However, $C^{\mathrm{DMFT}}_{RLRL'}$ is still local.
With this effective potential parameter, the LDA+DMFT level Green's function can be expressed 
in a similar form as the LDA-level Green's function, viz.,
\begin{multline}
[G^{\gamma,\mathrm{DMFT}}_{RLR'L'}({\bf k},z)]^{-1}=\\
\Delta^{1/2}_{Rl}[P^{\gamma, \mathrm{DMFT}}_{RLRL'}(z)-S^{\gamma}_{RLR'L'}({\bf k})]\Delta^{1/2}_{R'l'},
\end{multline}
where
\begin{align}
P^{\gamma, \mathrm{DMFT}}_{RL RL'}(z) &\equiv P^{\gamma,\mathrm{LDA}}_{Rl}(z)-\Delta^{-1/2}_{Rl}\Sigma_{RLRL'}(z)\Delta^{-1/2}_{R'l'}\nonumber\\
&=\Delta^{-1/2}_{Rl}(z-C_{Rl}-\Sigma_{RLRL'}(z))\Delta^{-1/2}_{R'l'}\nonumber\\
&=\Delta^{-1/2}_{Rl}(z-C^{\mathrm{DMFT}}_{RLRL'}(z))\Delta^{-1/2}_{R'l'}.
\end{align}
Note that due to the self-energy, the effective potential function now has off-diagonal elements.

\subsection{The combined CPA and DMFT loop}\label{sec_cpaloop}

For disorder calculations using the CPA, it is more convenient to use the tight-binding $\beta$-representation, since in this case only the potential functions (and not the structure constants) are random~\cite{gu.je.83,ku.dr.90,ab.ve.91,ab.sk.93} (see also Appendix~\ref{app_lmto}).
In this case, the representation-dependent potential parameter $V^{\beta}$ (Eq.~(\ref{betaparam})) will be modified accordingly:
\begin{equation}
V^{\beta,\mathrm{DMFT}}_{RLRL'}(z) \equiv C_{Rl}+\Sigma_{RLRL'}(z)-\frac{\Delta_{Rl}}{\gamma_{Rl}-\beta_{l}}.
\end{equation}
For LDA+DMFT calculations, this form of $V^{\beta,\mathrm{DMFT}}_{RLRL'}$ should be used for the potential function~(\ref{betapotf}),
the path operator $g^{\beta,\mathrm{DMFT}}_{RLR'L'}$, and the Green's function~(\ref{greenbeta}), in order for the 
Dyson equation~(\ref{dmftdyson}) to be fulfilled. Note that the transformations in Eqs.~(\ref{pathtrans}) and~(\ref{greenbeta}) are now no longer simply scaled as in the LDA case, but are matrix multiplications
due to the presence of off-diagonal terms in the self-energy.

In the following, all quantities will be on the dynamical mean-field level, and we suppress the common superscript ``DMFT'' for the coherent medium path operators $\tilde{g}$, the alloy component path operators $g^i$, the coherent potential functions $\tilde{P}$, and the alloy component potential functions $P^i$, which will be defined below. The representation-superscript $\beta$ is kept. An additional superscript $n$ appears to represent the iterative nature of the equations. The superscript $i$ refers to the index enumerating the alloy components at a certain site. We also introduce
the concentration of the respective alloy components $i$, at a site $R$, as $c^i_R$ ($0 \leqslant c^i_R \leqslant$ 1, $\sum_i c^i_R=1$). In this paper, the DMFT impurity problem is solved within the imaginary-axis Matsubara frequency representation, where the Matsubara frequencies
are defined as $i\omega _{\xi}=(2\xi+1)i\pi T$, where $\xi=0,\pm1,...$, and $T$ is the temperature. In the following,
we use the shorthand $i\omega=\{i\omega_0, ..., i\omega_\xi, ... \}$ to represent the set of all Matsubara frequencies.
Note that the DMFT impurity problem in principle also can be solved on the real axis, depending on
which numerical solver is used. The corresponding CPA equations will also hold for the real energy axis.

\begin{figure*}
\includegraphics[scale=1.1]{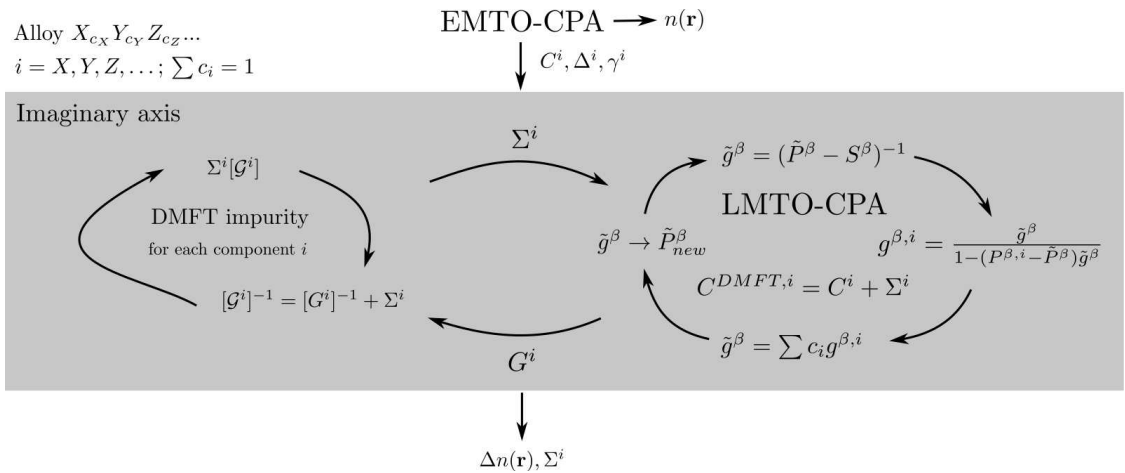}
\caption{Schematic flow diagram of the main DMFT-CPA loop, as implemented within the $z$MTO+DMFT
formalism~\cite{os.vi.17}. Inputs are the alloy component-dependent potential
parameters $C^i,\Delta^i$ and $\gamma^i$, taken from a LDA-level self-consistent EMTO-CPA calculation.
For each Matsubara frequency along the imaginary energy axis, the CPA equations are solved self-consistently.
The alloy impurity Green's functions are supplied to the DMFT impurity problem,
which is solved self-consistently, giving alloy component self-energies as output.
These self-energies in turn modify the potential parameters which is the key for the 
combined self consistency of CPA and DMFT equations.
To make the scheme charge 
self-consistent, the change in the density due to correlation, $\Delta n(\mathbf{r})$,
can be added to the EMTO-CPA charge density $n(\mathbf{r})$, and then the Kohn-Sham equations
are iterated until convergence.}\label{flow}
\end{figure*}

The CPA self-consistency condition requires that the sequential substitution
by an impurity atom into an effective, translationally-invariant, coherent
medium should produce no further electron scattering, on average.
In Appendix~\ref{app_cpadmft} we briefly review the CPA+DMFT algorithm for the case of
a one-band model system.
In the model Hamiltonian formalism~\cite{ve.ki.68,soven.67} the ensemble average
(coherent) Green's function ($G_c$) is constructed from the on-site
restricted averages of the component Green's function ($G^i$), according 
to the formula $G_c=\sum_i c^i_R G^i$. 
A similar formula can be written
for muffin-tin potentials in the multiple scattering formalism, following Gy\H{o}rffy~\cite{gyorffy.72}, and the averaging of the Green's function can be transfered   
to the path operator: 
\begin{equation}\label{cpaeq}
\tilde{g}^{\beta}_{RLRL'}(i\omega)=\sum\limits_{i} c^i_R g^{i,\beta}_{RLRL'}(i\omega).
\end{equation}
The coherent path operator in the $\beta$-representation,
\begin{equation}\label{cohpth}
\tilde{g}^{\beta}_{RLRL'}(i\omega)=\int [\tilde{P}^{\beta}(i\omega)-S^{\beta}({\bf k})]^{-1}_{RLRL'}\,\mathrm{d}{\bf k},
\end{equation}
has been integrated over the Brillouin zone (BZ). 
In Eq.~(\ref{cohpth}), the
coherent potential function $\tilde{P}^{\beta}(i\omega)$ has been introduced.
The alloy component
path operators in Eq.~(\ref{cpaeq}) are found through a Dyson equation,
\begin{multline}\label{cpadyson}
g^{i,\beta}_{RLRL'}(i\omega)=\tilde{g}^{\beta}_{RLRL'}(i\omega)+\sum\limits_{L''L'''}\tilde{g}^{\beta}_{RLRL''}(i\omega) \\
\times \left[P^{i,\beta}_{RL''RL'''}(i\omega)-\tilde{P}^{\beta}_{RL''RL'''}(i\omega)\right]g^{i,\beta}_{RL'''RL'}(i\omega).
\end{multline}
Here the potential functions $P^{i,\beta}(i\omega)$ are computed according to Eq.~(\ref{betapotf}),
for each type respectively. In order
to close the CPA equations self-consistently, a new coherent potential function has to be
determined at each iteration. This is done by taking the difference between the inverses
of the coherent path operators from the present iteration $n+1$ and the previous iteration $n$,
as follows:
\begin{equation}\label{updatep}
\tilde{P}^{\beta,n+1}(i\omega)=\tilde{P}^{\beta,n}(i\omega)-[\tilde{g}^{\beta,n+1}(i\omega)]^{-1}+[\tilde{g}^{\beta,n}(i\omega)]^{-1}.
\end{equation}
The new coherent potential function can be inserted into Eq.~(\ref{cohpth}), and the
cycle can be repeated until self-consistency has been reached. This is performed for each Matsubara frequency $i\omega$. Once self-consistency in the CPA equations has been achieved, the Green's functions for
each alloy component can be obtained by normalizing the alloy component path operators in Eq.~(\ref{cpadyson}),
using the transformation in Eq.~(\ref{greenbeta}). These (local) Green's functions are then used
as input for the DMFT impurity problem, with a separate impurity problem for each alloy component.
For a given interaction strength $U_{mm'm''m'''}$ (defined below)
on a particularly chosen alloy component (i.e. an atomic impurity embedded in the CPA coherent medium), 
we solve the interacting problem and alloy component self-energies 
are produced. The average over the disorder corresponds in 
this case to applying the CPA as described above in Eqs.~(\ref{cpaeq})-(\ref{updatep}). 
The coherent self-energy is an implicit quantity, in the self-consistency loops 
the alloy component self-energies and the path operators are used. 

The scheme presented above can easily be incorporated within the formalism of the $z$MTO+DMFT method~\cite{os.vi.17}.
Here, the EMTO method (see Appendix~\ref{app_emto} for a brief review) is employed to solve
the Kohn-Sham equations for random alloys self-consistently, within the CPA, on the level of the LDA.
The DMFT impurity problem is then solved in the Matsubara representation, using
linearization techniques to evaluate the alloy components Green's functions, as presented above. This can be done both
on the LDA-level (setting $\Sigma=0$) and on the DMFT-level (using the self-energy from the DMFT impurity problem).
The charge self-consistency is achieved similarly as in Ref.~[\onlinecite{os.vi.17}], by computing moments of the
alloy component Green's function at LDA and DMFT level.
The difference between the charge densities computed in this way can then be added as a correction on the LDA-level charge computed within the EMTO method.
In Figure~\ref{flow}, we present a schematic picture of the self-consistent loops.

\section{Results}\label{sec_res}

In order to demonstrate the feasibility of our proposed method, we apply it to investigate the electronic structure of the binary Cu$_{1-x}$Pd$_x$ alloy and the semi-Heusler compound
NiMnSb, with partially exchanged Ni and Mn components.

\subsection{Computational details}
\label{sec_comp_det}

In all calculations, the kink cancellation condition was set up for 16 energy points
distributed around a semi-circular contour with a diameter of 1 Ry, enclosing the valence band. The BZ integrations
were carried out on an equidistant $13 \times13 \times 13$ \textbf{k}-point mesh in the fcc BZ.
For the exchange-correlation potential the  local spin density approximation with the Perdew-Wang
parameterization~\cite{pe.wa.92} was used. For the studied alloys a $spd$-basis was used.
For the Cu-Pd system, the Cu $4s$ and $3d$ states, and for Pd the $5s$ and $4d$ states, were treated as valence.
For the case of NiMnSb, the Ni and Mn $4s$ and $3d$ states, and for Sb the $5s$ and $5p$ states,
were treated as valence.
The core electron levels were computed within the frozen-core approximation, and were treated
fully-relativistically. The valence electrons were treated within the scalar-relativistic approximation.
After self-consistency was achieved for NiMnSb, the density of states (DOS) was evaluated with a
$21 \times 21 \times 21$ \textbf{k}-point mesh, in order to get an accurate band gap.

To solve the DMFT equations, we used the spin-polarized
$T$-matrix fluctuation-exchange (SPTFLEX) solver~\cite{bi.sc.89,li.ka.98,ka.li.99,po.ka.05,gr.ma.12}. In this solver,
the electron-electron interaction term can be considered in a full spin and orbital rotationally invariant form, viz.
$\frac{1}{2}\sum_{{i \{m, \sigma \} }} U_{mm'm''m'''}
c^{\dag}_{im\sigma}c^{\dag}_{im'\sigma'}c_{im'''\sigma'}c_{im''\sigma} $.
Here, $c_{im\sigma}/c^\dagger_{im\sigma}$ annihilates/creates an electron with 
spin $\sigma$ on the orbital $m$ at the lattice site $i$.
The Coulomb matrix elements $U_{mm'm''m'''}$ are expressed in the usual
way~\cite{im.fu.98} in terms of Slater integrals.
In moderately correlated systems as studied here the modified multi-orbital 
fluctuation exchange (FLEX) approximation of Bickers and Scalapino~\cite{bi.sc.89} 
proved to be one of the most efficient approaches~\cite{li.ka.98,ka.li.99,po.ka.05,ka.ir.08}. 
The simplifications of the computational procedure in reformulating the 
FLEX as a DMFT impurity solver consists in neglecting dynamical interactions 
in the particle-particle channel, considering only static (of $T-$matrix type)
renormalization of the effective interactions. The fluctuating potential 
$W_{\sigma,-\sigma}(i\omega)$ is a complex energy-dependent matrix in spin 
space with off-diagonal elements 
$W^{\sigma,-\sigma}(i\omega)= U^{\tilde{m}} [ \chi^{\sigma,-\sigma}(i\omega) -\chi_0^{\sigma,-\sigma}(i\omega) ] U^{\tilde{m}}$,
where $U^{\tilde{m}}$ represents the bare vertex matrix corresponding
to the transverse magnetic channel, $\chi^{\sigma,-\sigma}(i\omega)$  
is an effective transverse susceptibility matrix, and $\chi_0^{\sigma,-\sigma}(i\omega)$ 
is the bare transverse susceptibility. The fermionic Matsubara
frequencies $i\omega$ were defined above and $\tilde{m}$ corresponds to 
the magnetic interaction channel\cite{ka.li.99,li.ka.98}. 
In this approximation the electronic self-energy
is calculated in terms of the effective interactions in various
channels. The particle-particle contribution to the self-energy
was combined with the Hartree-Fock and the second-order
contributions~\cite{ka.li.99,li.ka.98}, which adds to the 
the particle-hole contribution $\Sigma_{12\sigma}^{(ph)}=W^{\sigma,\sigma^\prime}_{1342}(i\omega)
G_{34}^{\sigma^\prime}(i\omega)$.
The local Green's functions as well as the electronic
self-energies are spin diagonal for collinear magnetic configurations.
Their pole structure, when analytically continued to the 
real energy axis, produce the appearance of the peaks located at specific 
energies determined by the materials characteristics and the 
symmetries of the orbitals.    

Since specific correlation effects are already
included in the exchange-correlation functional, so-called
``double counted'' terms must be subtracted. To achieve this, we replace
$\Sigma_{\sigma}(E)$ with $\Sigma_{\sigma}(E)-\Sigma_{\sigma}(0)$
\cite{li.ka.01} in all equations of the DMFT procedure \cite{ko.sa.06}.
Physically, this is related to the fact that DMFT only adds {\it dynamical}
correlations to the DFT result~\cite{pe.ma.03}.
The Matsubara frequencies were truncated after $\xi=$1024 frequencies, and the temperature
was set to $T=400$ K. The values for the average Coulomb $U$ and the exchange $J$ parameters are
discussed in connection with the presentation of the results in each case.
The densities of state were computed along a horizontal contour shifted
away from the real energy axis. At the end of the self-consistent calculations, to obtain the self-energy on the
horizontal contour, $\Sigma(i\omega)$ was analytically continued by a Pad\'e approximant~\cite{vi.se.77,os.ch.12}.

\subsection{Spectral functions and the Fermi surface of Cu$_{1-x}$Pd$_x$ random alloys}

\begin{figure*}
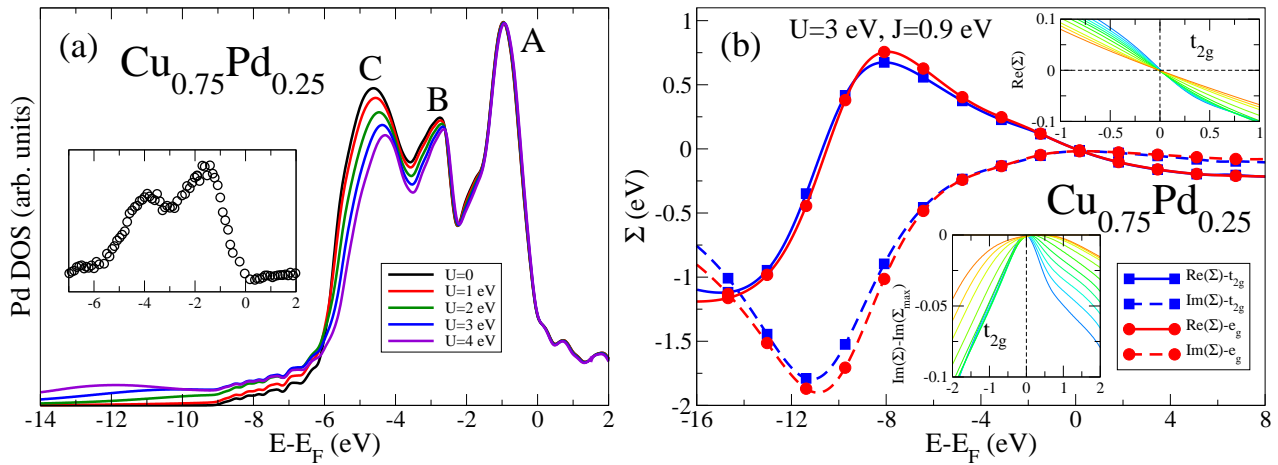

\includegraphics[scale=.35,clip=true]{fig2a.eps}
\includegraphics[scale=.35,clip=true]{fig2b.eps}
\caption{(a) Partial density of states of Pd in the Cu$_{0.75}$Pd$_{0.25}$ alloy, as a function of Coulomb interaction $U$. Inset: Experimental photoemission data taken from Ref.~[\onlinecite{wr.we.87}]. (b) Real parts (solid lines) and imaginary parts (dashed lines) of the Pd self-energy along the real-energy axis, for $U=3$ eV and $J=0.9$ eV. The $t_{2g}$-states are shown with blue boxes and the $e_g$-states with red disks. The insets in (b) are the real and imaginary parts of the $t_{2g}$ self-energy for different concentrations of palladium running from $x=1.0$ (blue) until $x=0.1$ (orange).} 
\label{cupddos}
\end{figure*}

We have previously investigated the electronic structure of fcc-Pd~\cite{os.ap.16},
within the framework of the LDA+DMFT method using the perturbative 
FLEX impurity solver~\cite{ka.li.02}. Recently, the properties of
fcc-Pd were revisited using a lattice (non-local) FLEX solver~\cite{sa.re.18}. These recent 
calculations~\cite{sa.re.18} support
our results using the local approximation of the self-energy.
Consequently, we study the electronic correlations in the CuPd alloys
using the same local DMFT technique as we used before. In particular we consider 
modeling correlations only for the Pd alloy component.

Discrepancies between the measured photoemission spectra~\cite{we.co.10}
and the KKR-CPA spectral functions~\cite{wr.we.87,wi.du.86} for various
Cu-Pd alloys were
often discussed in the literature. In particular LDA-CPA results for the 
Pd partial DOS of the Cu$_{0.75}$Pd$_{0.25}$ alloy reveal a 
three-peak structure (black line, Figure~\ref{cupddos}(a), peaks marked by A, B, and C), similar to
the DOS of pure fcc-Pd~\cite{os.ap.16}. 
Experimental data~\cite{wr.we.87} on the other hand, 
see also inset of Figure~\ref{cupddos}(a), show a contracted band width 
for the partial DOS and do not resolve the peak at the bottom of the
band (marked by C).
A detailed discussion concerning these discrepancies can be found in Ref.~[\onlinecite{we.co.10}]. We note that the frequently discussed
reasons for these discrepancies are connected to matrix element 
effects~\cite{na.mo.93}, broadening by electronic self-energy~\cite{na.mo.93}, 
and local lattice distortions~\cite{we.wr.87.2,th.un.94,ku.pe.98}, that go beyond the capabilities of 
standard CPA. 
Although it is not our intention to address all of the above inconsistencies,
our current implementation allows us to address the possible source of discrepancy in connection to the  
combined disorder and correlation effects. 

In Figure~\ref{cupddos}(a) we present the spectral function (DOS) for the Cu$_{0.75}$Pd$_{0.25}$ alloy, 
as a function of the Coulomb parameter $U$. All curves were evaluated
at the lattice constant given by a linear interpolation between that 
of pure Cu and pure Pd (Vegard's law), which
in this case corresponds to $a=3.68$ {\AA}. Vegard's law has previously been shown to hold
in a large range of concentrations 
for Cu-Pd within KKR-CPA~\cite{wi.hu.01}. As the Coulomb interaction is increased, 
the peak close to the bottom of the band (C) shifts towards the Fermi energy, while the major peak close to
$E_F$ (A) remains unchanged. 
The high binding energy peak (C) loses intensity with increasing $U$,
and the spectral weight is shifted to higher binding energy, where it
builds up a satellite structure. 
A similar behavior in the spectral weight shift was also found for 
pure Pd~\cite{os.ap.16}. 
In Figure~\ref{cupddos}(b) we show the Pd self-energies along the real-energy axis
for the interaction strength of $U=3$~eV and $J=0.9$~eV. Note that for different 
values of these parameters a qualitatively similar behavior of the self-energy
is obtained. This behavior is that of a typical 
Fermi-liquid frequently encountered in transition metals~\cite{gr.ma.07}:
a real part (solid lines) that show a negative slope at $E_F$ and the 
corresponding imaginary parts with parabolic energy dependence around $E_F$. 
For the entire concentration range, the Cu-Pd alloys therefore exhibit 
a Fermi-liquid behavior, as seen in the insets of Figure~\ref{cupddos}(b). 
In these calculations we kept the lattice constant fixed for all concentrations
to demonstrate genuine disorder effects, while disregarding volume 
effects.
Within the normal Fermi-liquid assumption, the renormalization constant
 that measures the discontinuity 
of the momentum distribution at $k=k_F$ can be computed as
$Z^{-1}=1-\partial \mathrm{Re} \Sigma(E)/\partial E|_{E=E_F} =m^*/m_{LDA}$. 
We see that by decreasing Pd concentration (color shifts from blue to orange), 
the absolute value of the slope of the real part of the self-energy decrease monotonically
and the electronic effective mass decreases. The effective masses are of a similar
magnitude as in previous studies on Pd~\cite{os.ap.16,sa.re.18}.

The results of the calculations including self-energy effects shown 
in Figure~\ref{cupddos}(a) bring the spectral function more
in line with experimental photoemission data~\cite{wr.we.87} (see also inset).
Since we neglect matrix element effects due to the photoemission process,
as well as local lattice relaxations, 
we do not make a quantitative statement concerning 
the differences between theory and experiment. 
However, our calculation shows that the proposed method which 
combines correlation and alloy disorder effects,
provides the correct trend in the spectral function.

\begin{figure}
\includegraphics[scale=.55]{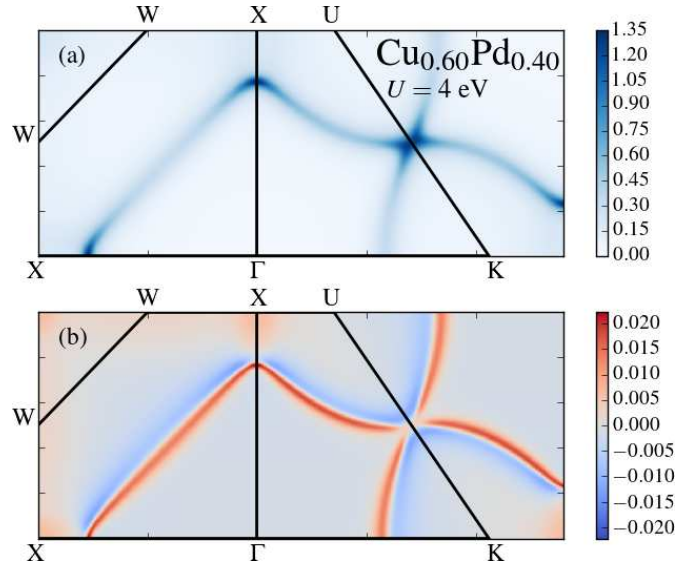}
\caption{(a) Fermi surface (Bloch spectral function) for the Cu$_{0.60}$Pd$_{0.40}$ alloy with $U$=4 eV.
(b) Difference of spectra between $U=4$ eV and $U=0$. The color maps indicate the spectral weight in arbitrary units.}
\label{cupd_bloch}
\end{figure}

In the following we comment upon the disorder and correlation 
induced modifications in the shape of the Fermi surface of CuPd alloys.
On the basis of KKR-CPA calculations Gy\H{o}rffy and Stocks~\cite{gy.st.83} proposed an electronic mechanism  
which determines short-range order effects experimentally 
seen in CuPd alloys. The experimental observation, namely the
dependence of the scattering intensities on concentration in these alloys,
was traced back to the flattening of the Fermi surface sheets
with increasing Pd concentration.
According to their results~\cite{gy.st.83} the Fermi surface must change from a convex shape
in the Cu-rich alloy to a concave one for the Pd-rich limit, in a continuous fashion.
Consequently the Fermi surface is forced to be almost flat for some concentration,
giving rise to nesting phenomena.
This was later confirmed by further experiments and CPA calculations~\cite{wi.hu.01,br.gi.01}.

\begin{figure*}
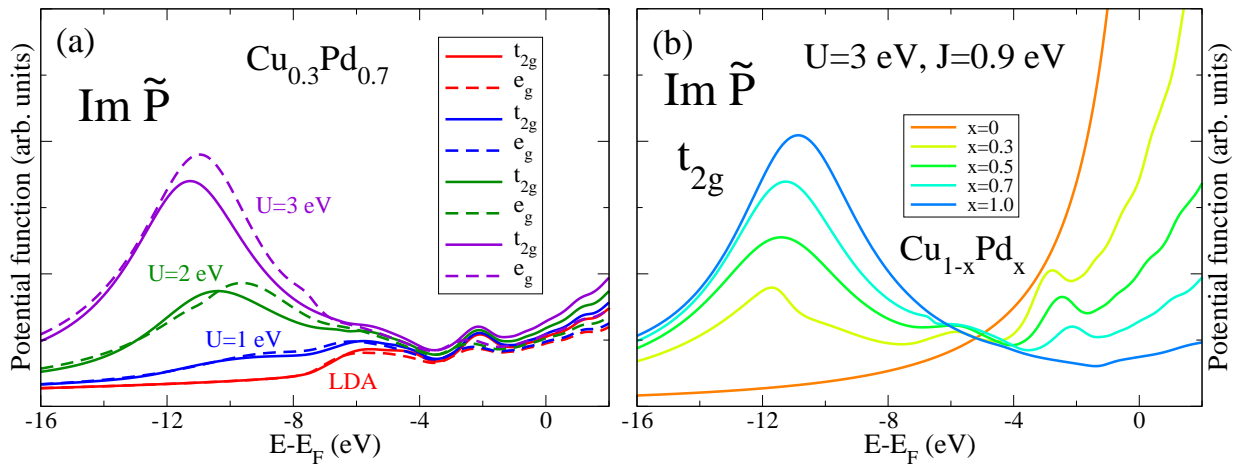

\includegraphics[scale=.35,clip=true]{fig4a.eps}
\includegraphics[scale=.35,clip=true]{fig4b.eps}
\caption{Imaginary part of the effective medium $\tilde{P}(E)$ along the real energy axis.
(a) $\tilde{P}(E)$ for several values of the Coulomb parameters and for a fixed concentration Cu$_{0.3}$Pd$_{0.7}$.
Red lines correspond to the LDA solution ($U=0$). $t_{2g}$ states are shown with solid lines
and $e_g$ states are shown with dashed lines. (b) $\tilde{P}(E)$ for several values of
the concentration $x$ and fixed Coulomb $U=3$~eV and Hund's coupling $J=0.9$~eV.
The concentration ranges from pure Cu ($x=0$, orange line) to pure Pd ($x=1$, blue line).
Only the $t_{2g}$-orbital is shown.}
\label{ptilde}
\end{figure*}

According to previous calculations~\cite{gy.st.83} a flattened Fermi surface in the $\Gamma X K$ plane was obtained
for the Cu$_{0.60}$Pd$_{0.40}$ alloy. 
In Figure~\ref{cupd_bloch}(a) we plot our results for the Fermi surface
of the same alloy. In our calculations we used for the 
lattice constant the value $a=3.72$~{\AA} (from Vegard's law), and electronic interactions 
on the Pd alloy component were parameterized by $U=4$~eV, and $J=1.2$~eV.
The Fermi surface is represented in the (010) and (110) planes of the fcc BZ.
The major part of the Fermi surface consists of the electron sheet centered at the $\Gamma$-point.
This sheet goes from convex to concave with Pd-alloying, forcing parts of the sheet to be nearly flat
at $40\%$ Pd. Our result is in good agreement with previous KKR-CPA calculations~\cite{wi.hu.01,br.gi.01}.
To quantify the effect of correlation, we plot in  Figure~\ref{cupd_bloch}(b)
the difference between the correlated ($U=4$ eV) and the non-correlated ($U=0$) case.
Note the relatively small scale, which shows 
that the Fermi surface is insensitive to correlation effects. This is expected due to
the Fermi-liquid behavior of the system.

The effective medium in the CPA theory 
is an auxiliary quantity that plays a similar role as the effective potential in the 
Kohn-Sham DFT.
Contrary to the effective Kohn-Sham DFT potential, the potential parameter $\tilde{P}_{RLRL}(E)$
representing the CPA effective medium is a local, complex, orbital, and energy dependent quantity. 
In the absence of electronic correlations ($U=0$) the potential parameter becomes diagonal in the orbital 
index. In addition, in the absence of disorder the orbital-resolved potential parameters 
reduce to the form seen in Eq.~(\ref{betapotf}), and were
shown to increase monotonically~\cite{gu.je.83}.
In Figure~\ref{ptilde}(a) we plot its imaginary part  
$\mathrm{Im} \tilde{P}_{RLRL}(E)$ for a given concentration (the case of 
Cu$_{0.3}$Pd$_{0.7}$) and for different values of the average Coulomb parameter $U$.
We note that at the LDA level in the presence of disorder ($x=0.3$), $\mathrm{Im} \tilde{P}(E)$ 
develops a similar structure to the density of states, which drops at about $-8$~eV,
corresponding to the bottom of the band. 
The connection between the DOS and the effective medium potential parameter is realized thorough the impurity Dyson equation Eq.~(\ref{cpadyson}). 
Increasing 
the $U$ values in the range of $1$ to $3$~eV we see a structure at higher binding energies,
related to the appearance and development of the additional pole structure of the 
fluctuating part of the dynamical self-energy described in Sec.~\ref{sec_comp_det}.
In Figure~\ref{ptilde}(b) we study the concentration dependence of $\mathrm{Im} \tilde{P}_{RLRL}(E)$
for the $t_{2g}$ orbitals, obtained for the fixed values of $U=3$~eV and
$J=0.9$~eV. 
For the pure case ($x=0$) no electronic correlations are considered and the energy dependence
of the potential parameter follows the LMTO description~\cite{gu.je.83,ande.75}.
The many-body effects, similar as seen in Fig.~\ref{ptilde}(a), gain in importance 
for larger concentrations and ultimately gives rise to
the satellite structure in the DOS (see Figure~\ref{cupddos}).

\subsection{Interplay of correlation and disorder in Mn-Ni partially interchanged NiMnSb}

\begin{figure*}
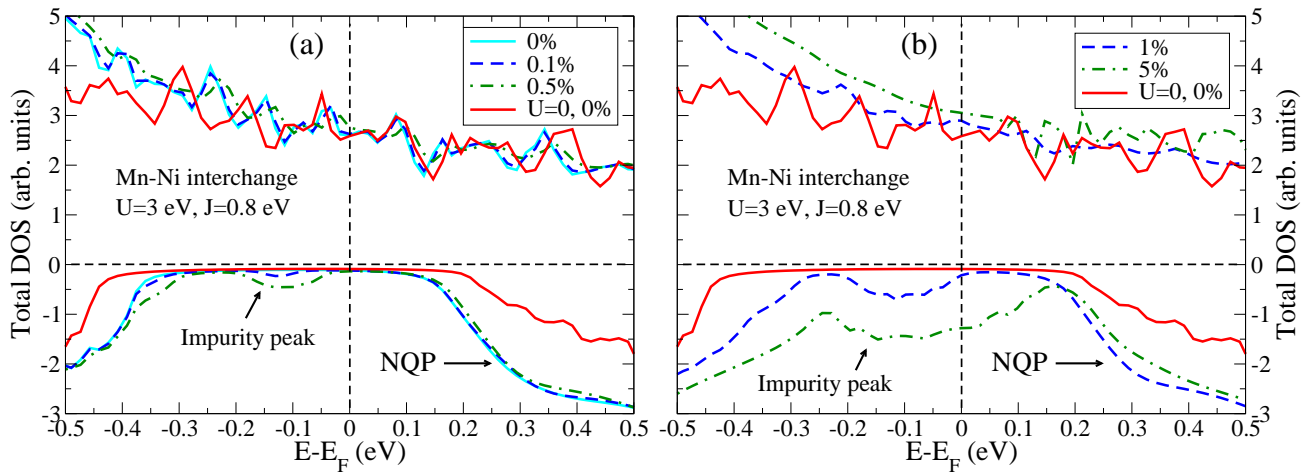

\includegraphics[scale=.35,clip=true]{fig5a.eps}
\includegraphics[scale=.35,clip=true]{fig5b.eps}
\caption{(a) Spin-resolved spectral function around the Fermi level for NiMnSb with Mn-Ni interchange,
from $0\%$ (no disorder, light blue solid line) to $0.1\%$ (dark blue dashed line),
$0.5\%$ (green dash-dotted line). The red solid line corresponds to $U=0$ and no disorder.
(b) Same as in (a), but for the larger degrees of disorder $1\%$ (dark blue dashed line),
and $5\%$ (green dash-dotted line).}\label{nimnsbdos}
\end{figure*}

Half-metallic ferromagnets (HMF)~\cite{ka.ir.08} are ferromagnetic systems
which are metallic in one spin channel, while for the opposite
spin direction the Fermi level is situated in a gap. Such systems 
 would therefore present a full spin-polarization at the Fermi level,
and have consequently drawn considerable interest due to
their potential application in spintronics.
One of the first systems to be characterized as a HMF is the semi-Heusler
NiMnSb~\cite{gr.mu.83}. 
The crystal structure of the NiMnSb compound is cubic with the space group $F\bar{4}3m$ (No. 216).
It consists of four interpenetrating fcc sublattices
equally spaced along the [111] direction. 
The Ni lattice sites are situated 
at $(0, 0, 0)$, Mn sites are at $(1/4, 1/4, 1/4)$, and Sb is situated at 
$(3/4, 3/4, 3/4)$. The position at $(1/2, 1/2, 1/2)$ is unoccupied in the
ordered alloy. 
In experiment, contrary to the DFT prediction, the
measured spin-polarization of NiMnSb is only $58\%$~\cite{so.by.98}.
Several suggestions have been given to explain this large reduction in
spin-polarization. Among them we mention electronic correlation effects~\cite{ch.ka.03,ka.ir.08} and disorder~\cite{or.fu.99,at.fa.04,ek.la.10}. 

Within the current implementation we have the opportunity to study the
combined effect at equal footing.
In the present calculations we use the experimental lattice constant, 
$a=5.927$~{\AA}. To parametrize the Coulomb interaction the values $U=3$ eV
and $J=0.8$ eV were used, which are in
the range of previous studies~\cite{ch.ka.03,al.ch.10,mo.ap.17}. Only the Mn $d$
states were
treated as correlated. Because the Ni $3d$ bands in NiMnSb are almost filled, these are subject to minor correlation effects, as shown previously~\cite{mo.ap.17}. 
The effect of electronic correlations is the appearance of nonquasiparticle (NQP) states
in the minority spin gap (spin down channel) just above the Fermi
level. The origin of these many-body NQP states is connected with
``spin-polaron'' processes: the spin-down low-energy electron excitations, which are forbidden for the HMF in the
one-particle picture, turn out to be allowed as superpositions of
spin-up electron excitations and virtual magnons~\cite{ed.he.73,ka.ir.08}. 
By direct computation, spin-orbit effects were found to be negligible~\cite{ma.sa.04.es} in NiMnSb. 
A partially filled minority spin gap was obtained but the material remains essentially half-metallic with a polarization of the DOS of about $99\%$~\cite{ma.sa.04.es}. The interplay of spin-orbit induced states and NQP states have been also discussed~\cite{ch.ar.06}. In contrast with the spin-orbit coupling, correlation induced NQP states
have a large asymmetric spectral weight in the minority-spin channel~\cite{ch.sa.08}, leading to a peculiar finite-temperature spin depolarization effect.
It has been shown that also disorder induces minority-spin states in
the energy gap of the ordered material~\cite{or.fu.99}. These states
widen with increasing disorder. This behavior leads to a reduced
minority-spin band gap and a shift of the Fermi energy
within the original band gap. 

We consider the partial interchange of Ni and Mn, (Ni$_{1-x}$Mn$_{x}$)(Mn$_{1-x}$Ni$_{x}$)Sb,
which leaves the overall stoichiometry and number of electrons constant.
In Figure~\ref{nimnsbdos} we show the total DOS around the Fermi level for 
different Mn-Ni interchange configurations and different concentrations. 
For the pure NiMnSb the LSDA minority occupied bonding states are mainly 
of Ni-$d$ character and are separated by a gap about 0.5 eV wide,
while unoccupied anti-bonding states are mainly of Mn-$d$ character~\cite{gr.mu.83,ka.ir.08,ya.ch.06}. 
It was pointed out in Ref.~\onlinecite{gr.mu.83} that the opening of a gap 
is assisted by Sb through symmetry lowering with the consequence that 
the distinction between Mn-$t_{2g}$ and Sb-$p$ character of the electrons 
is lost. In the majority spin channel (spin up), Ni-Mn covalency 
determines the presence of states at $E_F$ with dominant $d$-character.
The pure NiMnSb is ferromagnetic~\cite{gr.mu.83,ka.ir.08}, 
with a total ferromagnetic (integer) moment of 4 $\mu_B$, 
with the main contribution stemming from the Mn-site ($\sim 3.8$ $\mu_B$). 
Upon interchange disorder the Mn-moment at the Ni site is of opposite sign, 
$\sim -2$ $\mu_B$, while the Mn-moment at the Mn-site remains positive 
$\sim 3.7$ $\mu_B$. These values remain more or less unchanged by the 
presence of electronic correlations. The ferromagnetic states results from the
exchange interaction of Mn spins which are situated relatively far 
form each other~\cite{sa.sa.05,ka.ir.08}. The presence of interchange disorder 
induce magnetic moments of opposite sign on the neighboring (Ni$_{1-x}$Mn$_x$) and (Mn$_{1-x}$Ni$_x$) sites, as a consequence of magnetic couplings involving both direct and 
mediated exchange through Ni and Sb atoms. The larger the Mn-Ni interchange, the 
smaller the total magnetic moment.  

\begin{figure*}
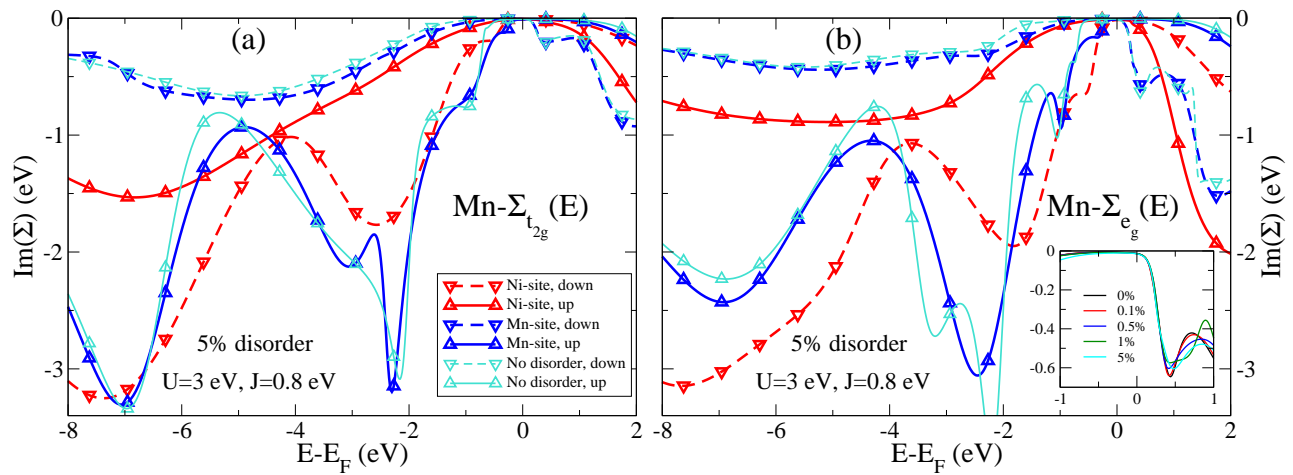

\includegraphics[scale=.35,clip=true]{fig6a.eps}
\includegraphics[scale=.35,clip=true]{fig6b.eps}
\caption{(a) Self-energy along the real axis for the $d-t_{2g}$ states of Mn, at a Mn-Ni interchange disorder of $5\%$. Dark blue lines correspond to Mn on the Mn-site, red lines correspond to Mn on the Ni-site. Dashed lines with triangles pointing down correspond to the spin-down channel, while solid lines with triangles pointing up correspond to the spin-up channel. The Mn self-energy for NiMnSb \emph{without} disorder (light blue) has been plotted for comparison. (b) Same as in (a), but for the Mn $d-e_g$ states. The inset in (b) shows the NQP states in the $e_g$ spin-down channel for different disorder concentrations.}\label{nimnsbsigma}
\end{figure*}

In the Figure~\ref{nimnsbdos}(a)/(b)
the LSDA(+DMFT) DOS for smaller degrees of disorder $x=0.1$, $0.5\%$ and
respectively for larger disorder $x=1\%$, $5\%$ is seen. The results 
for the clean, $x=0\%$ (ideal) case, NiMnSb, are presented with red lines
(non-interacting, $U=0$), and light blue lines (DMFT).  
Already at $0.1\%$ disorder (dark blue dashed line) minority states 
appear below $E_F$.
These states are generated by the presence of Ni impurities at the
Mn site, as previously shown by Orgassa \emph{et al.}~\cite{or.fu.99}.
Furthermore, the upper band edge is shifted to higher energy. As the disorder
is increased, the width of the Ni impurity states are increased.
With correlation, minority spin states appear just above the Fermi level. These
NQP states arise from many-body electron-magnon interactions~\cite{ka.ir.08}.
At larger degrees of disorder, see Figure~\ref{nimnsbdos}(b), the impurity states and the NQP states
overlap in energy, removing the spin-down gap. Hence, the combination of disorder 
due to the interchange between Ni and Mn sites 
and electronic correlation effects remove the half-metallic gap in NiMnSb.

Figure~\ref{nimnsbsigma}(a)/(b) displays the self-energy along the real energy axis for
the Mn $t_{2g}$/$e_g$ states, respectively, for
a Mn-Ni interchange of $5\%$. The dark blue lines correspond to the Mn at the Mn-site, and is similar to
the self-energy for pure NiMnSb (light blue lines). The self-energy behaves as in previous
calculations~\cite{ch.ka.03}, namely: the electrons within the spin-down channel (blue down-triangles) have 
a self-energy that is fairly small below $E_F$, but starts to increase above $E_F$.
At around $0.5$ eV above $E_F$, the self-energy shows a hump, which gives rise to the NQP
peak in the spectral function. The self-energy of spin-up electrons (blue up-triangles) behaves differently,
it is relatively large below $E_F$, while being small in magnitude above $E_F$.
The self-energy for the impurity Mn, situated at the Ni-site, is
marked by the red lines in Figure~\ref{nimnsbsigma}(a)/(b).
For the spin-down electrons (red down-triangle), the self-energy is large below $E_F$,
while it is small above $E_F$. The trend is opposite for the spin-up channel electrons (red up-triangle). 
This opposite behavior of the manganese self-energies at different sites  
reflects the anti-parallel configuration of the moments.

It is of interest to investigate how the effect of disorder, i.e. 
the degree of Mn-Ni interchange, influence the formation 
of NQP states. For this reason, in the inset of Figure~\ref{nimnsbsigma}(b), 
we plot the Mn-site self-energies of the 
$e_g$ orbitals for 
different disorder concentrations. For minor degrees of Mn-Ni interchange 
(up to $5\%$), the sudden increase in $\textrm{Im}\;\Sigma(E)$ just above $E_F$  
(dashed blue lines of Figure~\ref{nimnsbsigma}), signaling the departure 
from Fermi liquid behavior, remains unaffected. 
It should also be noted that the Mn self-energy at the Ni-site (Figure~\ref{nimnsbsigma}, 
red lines), follows the Fermi liquid (quasiparticle) behavior 
$ \textrm{Im}\;\Sigma (E) \propto (E-E_F)^2$.  
The Ni $d$ band in NiMnSb is almost 
fully occupied, leaving little possibility for magnons to be excited, therefore 
weak electron-magnon interaction exists in the Ni-sublattice and no NQP states above $E_F$ are visible in the density of states.

\section{Conclusion and Outlook}
\label{sec_conc}
In this paper we developed a calculation scheme 
within the framework of the density functional theory,
which allows one to study 
properties of disordered alloys including electronic correlation effects. 
We model disorder 
using the coherent potential approximation and include local but dynamic 
correlations through dynamical mean field theory. 
Similar to our previous implementation~\cite{os.vi.17}, the DFT-LDA Green's 
function is computed directly on the Matsubara contour. Simultaneously
the CPA is implemented within the LMTO formalism also in the Matsubara representation.  
Within the LMTO formalism the CPA effective medium is naturally encoded in the potential
function, which alone contains the necessary information about the atomic configuration
(assuming that a suitable screening representation is chosen). As shown in this paper, 
the simple parameterization of the potential function allows us to easily embed the 
self-energy into the standard LMTO potential parameters. Accordingly, the previously 
developed CPA schemes within the various muffin-tin approximations can then
be used with only minor changes.

We presented results of the electronic structure calculation for two disordered alloys: the Cu$_{1-x}$Pd$_x$ system, and the half-metallic NiMnSb semi-Heusler, in which correlations were considered for Pd and Mn alloy components respectively.
For the case of the binary CuPd system, we see that the inclusion of electronic correlation
improves the agreement with the experimental spectral functions for $x=0.25$. 
For a Pd concentration of $x=0.4$ the Fermi surface, which is well captured already
on the level of the LDA,
remains more or less unchanged as correlation effects are turned on.
In the second example, the partial exchange of Mn and Ni in NiMnSb was investigated, simultaneously with correlation effects. 
Already for low levels of disorder, impurity states appear below the Fermi level, 
while many-body induced nonquasiparticle states appear just above the Fermi level. 
For larger degree of interchange both these states contribute in closing the minority-spin gap.

In the future, the present method will be extended to compute total energies within the 
full-charge density technique~\cite{vi.ko.94}, making it possible to study the energetics
of anisotropic lattice distortions~\cite{vi.ab.01} in alloys. 
Since the present method is implemented on the imaginary axis, one can also consider to change the impurity solver to a continuous-time quantum Monte Carlo~\cite{gu.mi.11} algorithm. This
would allow to compute disordered strongly correlated system without any bias.
Another interesting venue is to
change the \emph{arithmetic} configuration average used in the CPA to the \emph{geometric}
average used in typical medium methods~\cite{te.zh.17,te.zh.18}. This will allow one to investigate
the effects of Anderson localization~\cite{ande.58} in realistic materials.

\section*{Acknowledgments}
We greatly benefited from discussions with O. K. Andersen and D. Vollhardt,
whose advice are gratefully acknowledged. 
A\"O is grateful for the discussion
with P. Weightman concerning the experiments on the Cu-Pd system.
A\"O also thanks W. H. Appelt, M. Sekania, and M. Dutschke for help with the graphics.
Financial support of the Deutsche Forschungsgemeinschaft through the Research Unit FOR 1346
and TRR80/F6 is gratefully acknowledged.
LV acknowledges financial support from the Swedish Research Council,
the Swedish Foundation for Strategic Research, the Swedish
Foundation for International Cooperation in Research and
Higher Education, and the Hungarian Scientific Research Fund (OTKA 84078).
We acknowledge computational resources provided by the Swedish National 
Infrastructure for Computing (SNIC) 
at the National Supercomputer Centre (NSC) in Link\"{o}ping.

\appendix
\section{Relations and formulas within the LMTO method}\label{app_lmto}
Within the nearly-orthogonal $\gamma$-representation, the potential function takes
the simple form
\begin{equation}\label{potf}
P^{\gamma}_{Rl}(z)=\frac{z-C_{Rl}}{\Delta_{Rl}}.
\end{equation}
An insertion of this form into Eq.~(\ref{pathop}), and comparing with Eq.~(\ref{green}),
one sees that 
\begin{equation}\label{gammagreen}
g^{\gamma}_{RL'RL'}({\bf k},z)=\sqrt{\Delta_{Rl}}G^{\gamma}_{RLR'L'}({\bf k},z)\sqrt{\Delta_{R'l'}},
\end{equation}
i.e., the Green's function is the normalized path operator.

In the case of a random alloy, the potential parameters $C_{Rl}$, $\Delta_{Rl}$ and $\gamma_{Rl}$
will be site-dependent random parameters. Hence, both the potential function $P^{\gamma}_{Rl}(z)$,
Eq.~(\ref{potf}), and the structure constants $S^{\gamma}$, will be random within the
$\gamma$-representation. 
This can be seen from the transformation in Eq.~(\ref{transform}) to the $\gamma$-representation,
\begin{equation}
S^\gamma_{RLR'L'} = [S^0(1-\gamma S^0)^{-1}]_{RLR'L'}.
\end{equation}
Since $\gamma$ is (disorder) potential dependent,
so is the structure constants.
To avoid this, it is useful to switch to the tight-binding
$\beta$-representation~\cite{an.je.84}, as has been pointed out previously~\cite{gu.je.83,ku.dr.90,ab.ve.91,ab.sk.93}.
Within the tight-binding $\beta$-representation, the potential function takes the form
\begin{equation}\label{betapotf}
P^{\beta}_{Rl}(z)=\frac{\Gamma^{\beta}_{Rl}}{V^{\beta}_{Rl}-z}+\frac{1}{\gamma_{Rl}-\beta_{l}},
\end{equation}
where here the representation-dependent potential parameters $V^{\beta}$ and $\Gamma^{\beta}$
are given by
\begin{equation}\label{betaparam}
V^{\beta}_{Rl}=C_{Rl}-\frac{\Delta_{Rl}}{\gamma_{Rl}-\beta_{l}}, \quad
\Gamma^{\beta}_{Rl}=\frac{\Delta_{Rl}}{(\gamma_{Rl}-\beta_{l})^2}.
\end{equation}
The $\beta_l$-parameters can be found tabulated in several sources~\cite{an.je.84,sk.ro.91},
and are independent of the (disorder) potential.
Therefore, the structure constants $S^{\beta}$ depend only on the geometry of the underlying lattice,
and only the potential function $P^{\beta}$ is random.
The path operator in the $\beta$-representation, $g^{\beta}_{RLR'L'}({\bf k},z)$, is given
similarly as in Eq.~(\ref{pathop}).
The following relation allows to transform path operators between 
different representations~\cite{an.je.84,an.pa.86}:
\begin{equation}\label{pathtrans}
g^{\beta}(z)=(\beta-\gamma)\frac{P^{\gamma}(z)}{P^{\beta}(z)}+\frac{P^{\gamma}(z)}{P^{\beta}(z)}g^{\gamma}(z)\frac{P^{\gamma}(z)}{P^{\beta}(z)},
\end{equation}
where we have omitted the indices for simplicity.
Using this transformation, and Eqs.~(\ref{potf}), (\ref{betapotf}), and (\ref{gammagreen}), the
Green's function can be obtained from $g^{\beta}_{RLR'L'}({\bf k},z)$ as~\cite{sk.ro.91}
\begin{multline}\label{greenbeta}
G^{\gamma}_{RLR'L'}({\bf k},z)=\\
\frac{1}{z-V^{\beta}_{Rl}}+
\frac{\sqrt{\Gamma_{Rl}}}{z-V^{\beta}_{Rl}}
g^{\beta}_{RLR'L'}({\bf k},z)
\frac{\sqrt{\Gamma_{R'l'}}}{z-V^{\beta}_{R'l'}}.
\end{multline}
Note that the transformations in Eqs.~(\ref{pathtrans}) and~(\ref{greenbeta}) are simply energy-dependent
scalings of the path operator, since the potential parameters and the potential functions are 
diagonal matrices.

We here briefly mention the accuracy of the presented expressions.
The formulas as written above give correct energies up to second order
in $(\epsilon-\epsilon_{\nu})$. A way to improve on this is by a variational procedure~\cite{an.sa.98},
which produces a new Hamiltonian, giving eigenvalues correct to third order.
Correspondingly, the substitution $z \rightarrow z + (z-\epsilon_{\nu})^3 p$ in Eq.~(\ref{greenbeta}) gives a third-order expression for the potential function~\cite{an.pa.86,an.je.86}.
Here, $p=\langle \dot{\phi}^\gamma \vert \dot{\phi}^\gamma \rangle$ is a (relatively small) potential parameter.
In order to compare the spectra arising from the different orders of LMTO's, we
investigated the DOS for various systems using either second or third order potential functions, and comparing the result with the DOS computed from the Hamiltonian through the spectral representation. We found that while at second order there was no difference between the DOS, for third order there were clear differences between the spectra. This
can be attributed to the false poles present
in the third-order potential function~\cite{an.sa.98}, since
the energy dependence is now not linear, but cubic. In practice, we found
that this lead to a loss of spectral weight in the Green's function of
Eq.~(\ref{greenbeta}), compared to the spectral representation.
Hence, we in this paper only consider second order potential functions in Eq.~(\ref{greenbeta}).

\section{Exact Muffin-Tin Orbitals method}\label{app_emto}

One choice of basis for the solution of the Kohn-Sham equation~(\ref{kosh}) is the \emph{energy-dependent}
exact muffin-tin orbitals~\cite{an.je.94.2,vito.01,vitos.10}, $\bar{\psi}$.
They are constructed as a sum of the so called partial waves, the solutions of the radial
equations within the spherical muffin-tins, and of the solutions in the interstitial region.
Using this basis, the Kohn-Sham eigenfunctions can be expressed as
\begin{equation}\label{emto}
\Psi_j({\bf r}) = \sum\limits_{RL} \bar\psi^a_{RL}({\epsilon_j,{\bf r}_R})v^a_{RL,j},
\end{equation}
where the superscript $a$ denotes the screening representation used in the EMTO theory~\cite{an.je.94.2,vito.01}.

The expansion coefficients, $v^a_{RL,j}$, are determined so the $\Psi_j({\bf r})$ is
a continuous and differentiable solution of Eq.~(\ref{kosh}) in all space. 
This leads to an energy-dependent secular equation,
$K^a_{RLR'L'}(\epsilon_j)v^a_{RL,j} = 0$,
where $K^a_{RLR'L'}$ is the so called kink matrix, viz.
\begin{equation}
K^{a}_{RLR'L'}({\bf k},z) \equiv
a\delta_{RR'}\delta_{LL'}D^{a}_{RL}(z) - aS^{a}_{RLR'L'}({\bf k},z).
\end{equation}
$D^{a}_{RL}(z)$ denotes the EMTO logarithmic derivative function
\cite{vi.sk.00,vito.01}, and $S^{a}_{RLR'L'}({\bf k},z)$ is the slope matrix \cite{an.sa.00}.
The energy dependence of the kink matrix and the secular equation
poses no difficulties, since the DFT problem can be solved by Green's function techniques
(see, for example, Ref.~\onlinecite{ze.de.82}). By defining the path operator $g^a_{RLR'L'}({\bf k},z)$ as the inverse of the kink matrix,
\begin{equation}
\sum_{R''L''}K^a_{R'L'R''L''}({\bf k},z)g^a_{R''L''RL}({\bf k},z) = \delta_{R'R} \delta_{L'L},
\end{equation}
the poles of the path operator in the complex energy plane will correspond to the eigenvalues
of the system. 
The energy derivative of the kink matrix, ${\dot K}_{RLR'L'}({\bf k},z)$,
gives the overlap matrix for the EMTO basis set~\cite{an.sa.00}, and hence it can be used to 
normalize the path operator $g_{R''L''RL}({\bf k},z)$, which gives the EMTO Green's function~\cite{vi.sk.00,vito.01}
\begin{eqnarray}\label{emtogreen}
G_{RLR'L'}({\bf k},z)\;&=&\;
\sum_{R''L''}g_{RLR''L''}({\bf k},z){\dot K}_{R''L''R'L'}({\bf k},z)\nonumber\\
&&-\delta_{RR'}\delta_{LL'} I_{RL}(z),
\end{eqnarray}
where $I_{RL}(z)$ accounts for the unphysical poles of ${\dot K}_{RLR'L'}(z)$~\cite{vito.01,vitos.10}.
The use of Green's functions also facilitates the implementation of the CPA, the reader
is referred to Refs.~\onlinecite{vito.01,vi.ab.01,vitos.10} for more detailed discussions.

\section{Illustrative CPA+DMFT algorithm for model calculations}\label{app_cpadmft}
The algorithm presented in Sec.~\ref{sec_cpaloop} is formulated in the 
language of multiple scattering for muffin-tin potentials. It represents a
generalization of the usual algorithm used for model calculations. 
In what follows we illustrate the CPA+DMFT self-consistency loop
for the one-band Hubbard model with on-site disorder
(or the so called Anderson-Hubbard Hamiltonian):
\begin{equation}\label{eq:c1}
H=-\sum_{ij\sigma} t_{ij} c^\dagger_{i\sigma} c^{}_{j\sigma}  + U\sum_i n_{i\uparrow}
n_{i\downarrow} + \sum_{i\sigma}(\epsilon_i-\mu_\sigma) n_{i\sigma}
\end{equation}
Here $c^\dagger_{i\sigma} (c^{}_{i\sigma})$ create (annihilate) a spin-$\sigma$ electron 
on site $i$ with $n_{i\sigma}=c^\dagger_{i\sigma} c^{}_{i\sigma}$ and $\mu_\sigma$
is the chemical potential of the spin-$\sigma$ electrons. The on-site energies 
$\epsilon_i$ are chosen as random, while the hopping elements $t_{ij}$ are
independent of randomness; thus, short range order is neglected.
The coherent and alloy component Green's functions, and the 
corresponding self-energies, are complex functions of the real
energy $E$.  For a multi-band case these quantities are matrices 
in orbital space.  We start the self-consistency loop with a guess
for the self-energy $\Sigma_c$, which includes disorder and correlation effects. 
We emphasize that the combined disorder and correlation effects enter in one 
single self-energy $\Sigma_c$. 
The local Green's function $G_c=\sum_{\bf k}\left[  E+\mu - \epsilon({\bf k}) -\Sigma_c \right]^{-1}$ is computed
from the electronic dispersion $\epsilon({\bf k})$ (eigenstate of the lattice
Hamiltonian in the absence of disorder and electronic correlations) and the 
initial guess for the self-energy $\Sigma_c$. 
 From the coherent (local) Green's function, 
alloy component ($i=A,B,\dots$) Green's functions are computed as 
\begin{equation}\label{eq:c2}
G_{i} = G_c \left[1-(V_{i}+\Sigma^{DMFT}_i-\Sigma_c) G_c \right]^{-1},
\end{equation}
for a given (fixed) disorder realization. 
In the next step the many-body problem is solved using the DMFT methodology:
the DMFT bath Green's function is constructed as: $\mathcal{G}_i^{-1}=G_{i}^{-1}+\Sigma^{DMFT}_i$. Specific DMFT impurity solvers produce alloy component
many-body self-energies $\Sigma^{DMFT}_i[\mathcal{G}_{i}]$. In the next step we 
request that the alloy components should fulfill the 
CPA equation: $G_c=\sum_i c_i G_i$. This corresponds to the averaging over 
the disorder realizations. From the newly computed $G_c$
the coherent self-energy $\Sigma_c$ follows directly. 
To close the self-consistency loop $G_c$ and $\Sigma_c$ are returned into
the Eq.~(\ref{eq:c2}) to produce new alloy component Green's functions.
On a more formal level this algorithm was presented in Ref.~\onlinecite{ul.ja.95}.

Finally, we mention here the formal equivalence between the equations 
discussed in the present appendix with those shown in Sec.~\ref{sec_cpaloop}.
The CPA equation $G_c=\sum_i c_i G_i$ and Eq.~(\ref{cpaeq}) are equivalent. 
The local Green's function formula ($G_c=\sum_{\bf k}\left[  E+\mu - \epsilon({\bf k}) -\Sigma_c \right]^{-1}$) corresponds to the Eq.~(\ref{cohpth})
in the language of multiple scattering. Finally, 
Eq.~(\ref{eq:c2}) and Eq.~(\ref{cpadyson}) are equivalent as they provide the alloy components
computed using the Dyson equation.
In our recent paper~\cite{te.zh.17} we have extensively discussed several 
self-consistent loop algorithms for the disorder problem. These include
cluster extensions and alternative effective medium theories beyond the CPA.

\bibliography{references_database,references_cpa}
\end{document}